\documentclass[journal]{IEEEtran}

\ifCLASSINFOpdf
  \usepackage[pdftex]{graphicx}
  \usepackage[table]{xcolor}
  \usepackage{hyperref}
    \usepackage{color}
    \usepackage{subcaption}
\else
\fi

\begin{document}
%

\title{FluentNet: End-to-End Detection of Speech Disfluency with Deep Learning}


%
%
%

\author{Tedd~Kourkounakis,
        Amirhossein~Hajavi,
        and~Ali~Etemad
\thanks{All the authors are with the Department
of Electrical and Computer Engineering, Queen's University, Kingston,
ON, K7L3N6 Canada, e-mail: tedd.kourkounakis@queensu.ca.}}


%

\maketitle

\begin{abstract}
Strong presentation skills are valuable and sought-after in workplace and classroom environments alike. 
Of the possible improvements to vocal presentations, disfluencies and stutters in particular remain one of the most common and prominent factors of someone's demonstration. Millions of people are affected by stuttering and other speech disfluencies, with the majority of the world having experienced mild stutters while communicating under stressful conditions. While there has been much research in the field of automatic speech recognition and language models, there lacks the sufficient body of work when it comes to disfluency detection and recognition. To this end, we propose an end-to-end deep neural network, FluentNet, capable of detecting a number of different disfluency types. FluentNet consists of a Squeeze-and-Excitation Residual convolutional neural network which facilitate the learning of strong spectral frame-level representations, followed by a set of bidirectional long short-term memory layers that aid in learning effective temporal relationships. Lastly, FluentNet uses an attention mechanism to focus on the important parts of speech to obtain a better performance. We perform a number of different experiments, comparisons, and ablation studies to evaluate our model. Our model achieves state-of-the-art results by outperforming other solutions in the field on the publicly available UCLASS dataset. Additionally, we present LibriStutter: a disfluency dataset based on the public LibriSpeech dataset with synthesized stutters. We also evaluate FluentNet on this dataset, showing the strong performance of our model versus a number of benchmark techniques.

\end{abstract}


%
\IEEEpeerreviewmaketitle

\section{Introduction} \label{Introduction}

\IEEEPARstart{C}LEAR and comprehensive speech is the vital backbone to strong communication and presentation skills \cite{ferguson2018talker}. Where some occupations consist mainly of presenting, most careers require and thrive from the ability to communicate effectively. Research has shown that oral communication remains one of the more employable skills in both the perception of employers and new graduates \cite{robinson2007}. Simple changes to ones speaking patterns such as volume or appearance of disfluencies can have a huge impact on the ability to convey information effectively. By providing simplified, quantifiable data concerning ones speech patterns, as well as feedback on how to change ones speaking habits, drastic improvements could be made to anyone's communication skills \cite{mayo_clinic_stuttering}.  

In regard to presentation skills, disfluent speech remains one of the more common factors \cite{ROBOCOP}. Any abnormality or generally uncommon component of one's speech patterns is referred to as a speech disfluency \cite{ASHA}. There are hundreds of different speech disfluencies often grouped together alongside language and swallowing disorders. Of these afflictions, stuttering proves to be one of the most common and most recognized of the lot \cite{ASHA}.

Stuttering, also known as stammering, as a disorder can be generally defined as issues pertaining to the consistency of the flow and fluency of speech. This often involves involuntary additions of sounds and words, and the delay or inability to consistently progress through a phrase. 
Although labelled as a disorder, stuttering can occur in any person’s speech, often induced by stress or nervousness \cite{nhs}. These cases however do not correlate with stammering as a disorder, but are caused by performance anxiety \cite{adaa}. The use of stutter detection does not only apply to those with long term stutter impairments, but can appeal to the majority of the world as it can help with the improvement of communication skills. 

As the breadth of applications using machine learning techniques have flourished in recent decades, they have only recently began to be utilized in the field of speech disfluency and disorder detection. 
While deep learning has dominated many areas of speech processing, for instance speech recognition \cite{wang2020} \cite{he2019}, speaker recognition \cite{amir} \cite{snyder2019}, and speech synthesis \cite{prenger2019} \cite{li2019}, very little work has been done toward the problem of speech disfluency detection.
Disfluencies, including stutters, are not easily definable; they come in many shapes and variations. This means that factors such as gender, age, accent, and the language themselves will affect the contents of each stutter, greatly complicating the problem space. As well, there are many classes of stutter, each with their own sub-classes and with wildly different structures, making the identification of all stutter types with a single model a difficult task. Even a specific type of stutter applied to a single word can be conducted in a wide variety of ways. Where people are great at identifying stutters through their experience with them, machine learning models have historically struggled with this (as we show in Section \ref{Related Work}).

Another common issue is the sheer lack of sufficient training data available. Many previous works often rely on their own manually recorded, transcribed, and labelled datasets, which are often quite small due to the work involved in their creation \cite{howell1995} \cite{howell1997} \cite{dash2018} \cite{interspeech2018}. There is only one commonly used public dataset, UCLASS \cite{UCLASS}, that is widely used amongst works in this area, though it still is also quite small. 

Many disfluency detection solutions provide some form of filler word identification, flagging and counting any spoken interjections (e.g. `okay', `right', etc.). However, upon further investigation, these applications simply request a list of interjections from the user and use Speech-to-Text (STT) tools in order to match the spoken word with any interjections in the list. 
Though this may work fine for interjections such as \textit{`um'} and \textit{`uh'} (assuming the used STT tool has the necessary embeddings), this can lead to serious overall errors in classification for most other utterances that are actual words, such as \textit{`like'}, which is commonly used as a filler word in the English language. 

Early works in stutter detection, realizing the challenges mentioned above, first sought out to test the viability of identifying stutters from clean speech. These models primarily focused on machine learning models with very small datasets, consisting of a single stutter type, or even a single word \cite{howell1995}, \cite{tan2007}. In more recent years, and due to the rise of automatic speech recognition (ASR), language models have been used to tackle stutter recognition. These works have proven to be strong at identifying certain stutter types, and have been showing ever improving results \cite{interspeech2018}, \cite{dash2018}. However, due to the uncertainty surrounding  relations between cleanly spoken and stuttered word embeddings, it remains difficult for these models to generalize across multiple stutter types. It is hypothesized that by bypassing the use of language models, and by focusing solely on phonetics through the use of convolution networks, a model can be created that both maintains a strong average accuracy while also being effective across all stutter types.

In this paper, 
we propose a model capable of detecting speech disfluencies. To this end, we design FluentNet, a deep neural network (DNN) for automated speech disfluency detection. The proposed network does not apply any language model aspects, but instead focuses on the direct classification of speech signals. This allows for the avoidance of complex and time consuming ASR as a pre-processing steps in our model, and would provide the ability to view the scenario as an end-to-end solution using a single deep neural network. We validate our model on a commonly used benchmark dataset UCLASS \cite{UCLASS}. To tackle the issue of scarce stutter-related speech datasets, we also develop a synthetic dataset based on a non-stuttered speech dataset (LibriSpeech \cite{LibriSpeech}), which we entitle LibriStutter. This dataset is created to mimic stuttered speech and vastly expand the amount of data available for use. Our end-to-end neural network takes spectrogram feature images as inputs, and uses Squeeze-and-Excitation residual (SE-ResNet) blocks for learning the speech embedding. Next, a bidirectional long short-term memory (BLSTM) network is used to learn the temporal relationships, followed by an attention mechanism to focus on the more salient parts of the speech. Experiments show the effectiveness of our approach in generalizing across multiple classes of stutters while maintaining a high accuracy and strong consistency between classes on both datasets. 

The key contributions of our work can be summarized as follows:
(1) We propose FluentNet, an end-to-end deep neural network capable of detection of several types of speech disfluencies;
(2) We develop a synthesized disfluency dataset called \textit{LibriStutter} based on the publicly available LibriSpeech dataset by artificially generating several types of disfluencies, namely sound, word, and phrase repetitions, as well as prolongations and interjections. The dataset contains the output labels that can be used in training deep learning methods;
(3) We evaluate our model (FluentNet) on two datasets, \textit{UCLASS} and LibriStutter. The experiments show that our model achieves state-of-the-art results on both datasets outperforming a number of other baselines as well as previously published work;
(4) We make our annotations on the existing UCLASS dataset, along with the entire LibriStutter dataset and its labels, publicly available\footnote{1 \url{http://aiimlab.com/resources.html}} to contribute to the field and facilitate further research.

This is an extension of our earlier work titled ``Detecting Multiple Speech Disfluencies using a Deep Residual Network with Bidirectional Long Short-Term Memory'', published in the 2020 IEEE International Conference on Acoustics, Speech, and Signal Processing (ICASSP). This paper focused on tackling the problem of detection and classification of different forms of stutters. 
The model used a deep residual network and bidirectional long short-term memory layers to classify different types of stutters.
In this extended work, we replace the previously used residual blocks of the spectral encoder with residual squeeze-and-excitation blocks. Additionally, we add an attention mechanism after the recurrent network to better focus the network on salient parts of input speech. Furthermore, we develop a new dataset, which we present in this paper and make publicly available. Lastly, we perform thorough experiments, for instance through additional benchmark comparisons and ablation studies. Our experiments show the improvements made by FluentNet over our preliminary work, as validated on both the UCLASS dataset (previously used) as well as the newly developed dataset. This new model provides greater advancement towards end-to-end disfluency detection and classification.
The rest of this paper is organized as follows; a discussion of previous contributions towards stutter recognition in Section \ref{Related Work} followed by our methodology including a breakdown of the model in Section \ref{Proposed Method}, the datasets and benchmark models applied in Section \ref{Experiments}, a discussion of our results in Section \ref{Results}, and our conclusion in the final section.

\section{Related Work} \label{Related Work}
There has recently been increasing interest in the fields of deep learning, speech, and audio processing. However, as discussed earlier in section \ref{Introduction}, there has been minimal research targeting automated detection of speech disfluencies including stuttering, most likely  as a result of insufficient data and smaller number of potential applications in comparison to other speech-related problems such as speech recognition \cite{zeghidour2018} \cite{he2019} and speaker recognition \cite{amir} \cite{snyder2019}. In the following sections we first provide a summary of the type of disfluencies commonly targeted in the area, followed by a review of the existing work that fall under the umbrella of speech disfluency detection and classification.



\begin{table*}
\begin{center}
\small
\caption{Types of stutters considered for training and testing labels.}
\label{table: stutters}
\begin{tabular}{l l l l} 
\hline
Label & Stutter Disfluency & Description & Example \\
\hline\hline

S & Sound Repetition & Repetition of a phoneme & th-th-this \\
PW & Part-Word Repetition & Repetition of a syllable & bec-because \\
W & Word Repetition & Repetition of any word & why why \\
PH & Phrase Repetition & Repetition of multiple successive words &  I know I know that\\
R & Revision & Repetition of thought, rephrased mid sentence & I think that- I believe that \\
I & Interjection & Addition of fabricated words or sounds & um, uh, like\\ 
PR & Prolongation & Prolonged sounds & whoooooo is there \\
B & Block & Involuntary pause within a phrase & I want \textit{(pause)} to \\
\hline
\end{tabular}
\end{center}
\end{table*}

\subsection{Background: Types of Speech Disfluency}

There are a number of different stuttering types, often categorized into four main groups: repetitions, prolongations, interjections, and blocks. A summary of all these disfluency types and examples of each have been presented in Table \ref{table: stutters}. The descriptions for each of these categories is as follows.

Repetitions
are classified as any part of an utterance repeated at quick pace. As this definition still remains general, repetitions are often further sub-categorized \cite{ASHA}. These sub-categories have been used in previous works classifying stutter disfluencies \cite{yairiambrose} \cite{justeandrade} \cite{interspeech2018}, which include sound, word, and phrase repetitions, as well as revisions.
Sound repetitions (S) are repetitions of a single phoneme, or short sound, often consisting of a single letter. Part-word, or syllable repetitions (PW), as its name suggests, are classified as the repetition of syllables, which can consist of multiple phonemes. Similarly, word repetitions (W) are defined as any repetition of a single word, and phrase repetitions (PH) are the repetition of phrases, consisting of multiple consecutive words. The final repetition-type disfluency is revision (R). Similar to phrase repetitions, they consist of repeated phrases, where the repeated segment is rephrased, containing new or different information from the first iteration. 
A rise in pitch may accompany this disfluency type \cite{stutteringfoundation_pitch}.

Interjections (I), often referred to as filler words, consist of the addition of any utterance that does not logically belong in the spoken phrase. Common interjections in the English language include exclamations, such as \textit{`um'} and \textit{`uh'}, as well as discourse markers such as \textit{`like'}, \textit{`okay'}, and \textit{`right'}. 

Prolongation (PR) stutters are presented as a lengthened or sustained phoneme. The duration of these prolonged utterances vary alongside the severity of the stutter. Similar to repetitions, this disfluency is often accompanied by a rise in pitch.

The final category of stuttering are silent blocks (B), which are sudden cutoffs of vocal utterances. These are often involuntary and are presented as pauses within a given phrase.

\subsection{Stutter Recognition with Classical Machine Learning}

Before the focus of stutter recognition targeted maximizing accuracy in classification of stammers, a number of works were performed toward testing the viability of stutter detection. In 1995, Howell et al. \cite{howell1995}, who later helped to create the UCLASS dataset \cite{UCLASS} used in this paper, employed a set of pre-defined words to identify repetition and prolongation stutters. From these, they extracted the autocorrelation features, spectral information, and envelope parameters from the audio. Each was used as an input to a fully connected artificial neural network (ANN). Findings showed that the model achieved its strongest classification results against severe disfluencies, and was weakest for mild ones. These models were able to achieve a maximum detection rate of 0.82 on severe prolongation stutters.
Howell et al. \cite{howell1997} later furthered their work using a larger set of data, as well as a wider variety of audio parameters. This work also introduced an ANN model for both repetition and prolongation types, and more judges were used to identify stutters with strict restrictions towards agreement of disfluency labeling. Results showed that the best parameters for disfluency classification were fragmentation spectral measures for whole words, as well as duration and supralexical disfluencies of energy in part-words. 

Tan et al. \cite{tan2007} worked on testing the viability of stutter detection through a simplified approach in order to maximize the possible results. By collecting audio samples of clean, stuttered, and artificially generated copies of single pre-chosen words, they were able to reach an average accuracy of 96\% on the human samples using a hidden Markov model (HMM). This served as a temporary benchmark towards the possible best average results for stutter detection.

Ravikumar et al. have attempted a variety of classifiers on syllable repetitions, including an HMM \cite{ravikumarHMM} and support vector machine (SVM) \cite{ravikumar2009} using Mel-frequency cepstral coefficients (MFCCs) features. Their best results were obtained when classifying this stutter type using the SVM on 15 participants, achieved an accuracy of 94.35\%. No other disfluency types were considered.


A detailed summary of previously attempted stutter classification methods, including some of the aforementioned classical models, is available in the form of a review paper in \cite{chee2009_overview}. This paper provides background on the use of three different models (ANNs, HMMs and SVM) towards the application of stutter recognition. Of the works considered in that review paper in 2009, it was concluded that HMMs achieve the best results in stutter recognition.

\begin{table*}
\begin{center}
\scriptsize
\setlength\tabcolsep{2pt}
\caption{Summary of previous stutter disfluency classification methods.}
\label{table: literature}
\resizebox{\linewidth}{!}{\begin{tabular}{l l l p{4cm} p{4cm} l}
\hline
Year & Author & Dataset & Features & Classification Method & Results\\
\hline\hline 
1995 & Howell et al. \cite{howell1995} & N/A & autocorrelation function, spectral information, envelope parameters & ANN & Acc.: 82\%\\\hline
1997 & Howell et al. \cite{howell1997} & 12 Speech Samples & oscillographic and spectrographic parameters & ANN & Avg. Acc.: 92\%\\\hline
2007 & Tan et al. \cite{tan2007} & 20 Samples (single word) & MFCC & HMM & Acc.: 96\% \\\hline
2009 & Ravikumar et al. \cite{ravikumar2009} & 15 Speech Samples & MFCC & SVM & Acc.: 94.35\%\\\hline
2016 & Zayats et al. \cite{zayats2016} & Switchboard Corpus & MFCC & BLSTM w/ Attention & F1: 85.9\\\hline
2018 & Alharbi et al. \cite{interspeech2018} & UCLASS & Word Lattice & Finite State Transducer, Amplitude and Time Thresholding & Avg. MR: 37\%\\\hline
2018 & Dash et al. \cite{dash2018} & 60 Speech Samples & Amplitude & STT, Amplitude Thresholding & Acc.: 86\%\\\hline
2019 & Villegas et al. \cite{villegas2019} & 68 Participants & Respiratory Biosignals & Perceptron & Acc.: 95.4\%\\\hline
2019 & Santoso et al. \cite{santoso2019} & PASD, UUDB & MFCC & BLSTM w/ Attention & F1: 69.1\\\hline
2020 & Chen et al. \cite{Chen2020} & In-house Chinese Corpus & Word \& Position Embeddings & CT-Transformer & MR: 38.5\%\\
\hline
\end{tabular}}
\end{center}
\end{table*}

\subsection{Stutter Recognition with Deep Learning}
With the recent advancements in deep learning, disfluency detection and classification has seen an increase in popularity within the field with a higher tendency towards end-to-end approaches. ASR has become an increasingly popular method of tackling the problem of disfluency classification. As some stuttered speech results in repeated words, as well as prolonged utterances, these can be represented by word embeddings and sound amplitude features, respectively. To exploit this concept, Alharbi et al. \cite{interspeech2018} detected sound and word repetitions, as well as revision disfluencies using task-oriented finite state transducer (FST) lattices. They also utilized amplitude thresholding techniques to detect prolongations in speech. These methods resulted in an average 37\% miss rate across the 4 different types of disfluencies. 

Dash et al. \cite{dash2018} have used an STT model in order to identify word and phrase repetitions within stuttered speech. To detect prolongation stutters, they integrated a neural network capable of finding optimal cutoff amplitudes for a given speaker to expand upon simple thresholding methods. As these ASR works required full word embeddings to classify repetitions, they either fared poorly against, or did not attempt sound or part word repetitions.

Deep recurrent neural networks (RNN), namely BLSTM, have been used to tackle stutter classification. Zayats et al. \cite{zayats2016} trained a BLSTM with Integer Linear Programming (ILP) \cite{ILP} on a set of MFCC features to detect repetitions with an F-score of 85.9. Similarly, a work done by Santoso et al. applied a BLSTM followed by an attention mechanism to perform stutter detection based on input MFCC features, obtaining an maximum F-score of 69.1 \cite{santoso2019}. More recently in a study by Chen et al., a Controllable Time-delay Transformer (CT-Transformer) has been used to detect speech disfluencies and correct punctuation in real time \cite{Chen2020}. In our initial work on stutter classification, we utilized spectrogram features of stuttered audio and used a BLSTM \cite{kourkounakis2020} to learn temporal relationships following spectral frame-level representation learning by a ResNet. This model achieved a 91.15\% average accuracy across six different stutter categories.

In an interesting recent work, Villegas et al. utilized respiratory biosignals in order to better detect stutters \cite{villegas2019}. By correlating respiratory volume and flow, as well as heart rate measurements correlating to the time when a stutter occurs, they were able to classify block stutters with an accuracy of 95.4\% using an MLP. 

A 2018 summary and comparison of different features and classification methods for stuttering has been conducted by Khara et al. \cite{khara2018}. This work discusses and compares different popular feature extraction methods, classifiers and their uses, as well as their advantages and shortcomings. The paper discusses that MFCC feature extraction has historically provided the strongest results.
Similarly, ANNs provide the most flexibility and adaptability compared to other models, especially linear ones.

Table \ref{table: literature} provides a summary of the related works on disfluency detection and classification. It can be observed and concluded that disfluency classification has been progressing in one of two fronts \textit{i}) end-to-end speech-based methods, or \textit{ii}) language-based models relying on an ASR pre-processing step. Our work in this paper is positioned in the first category in order to avoid the reliance on an ASR step. Moreover, from Table \ref{table: literature}, we observe that although progresses is being made in the area of speech disfluency recognition, the lack of available data remains a hindrance to potential further achievements in the field.

\section{Proposed Method} \label{Proposed Method}
\subsection{Problem and Solution Overview}

\begin{figure}[!t]
    \begin{center}
    \includegraphics[width=0.7\columnwidth]{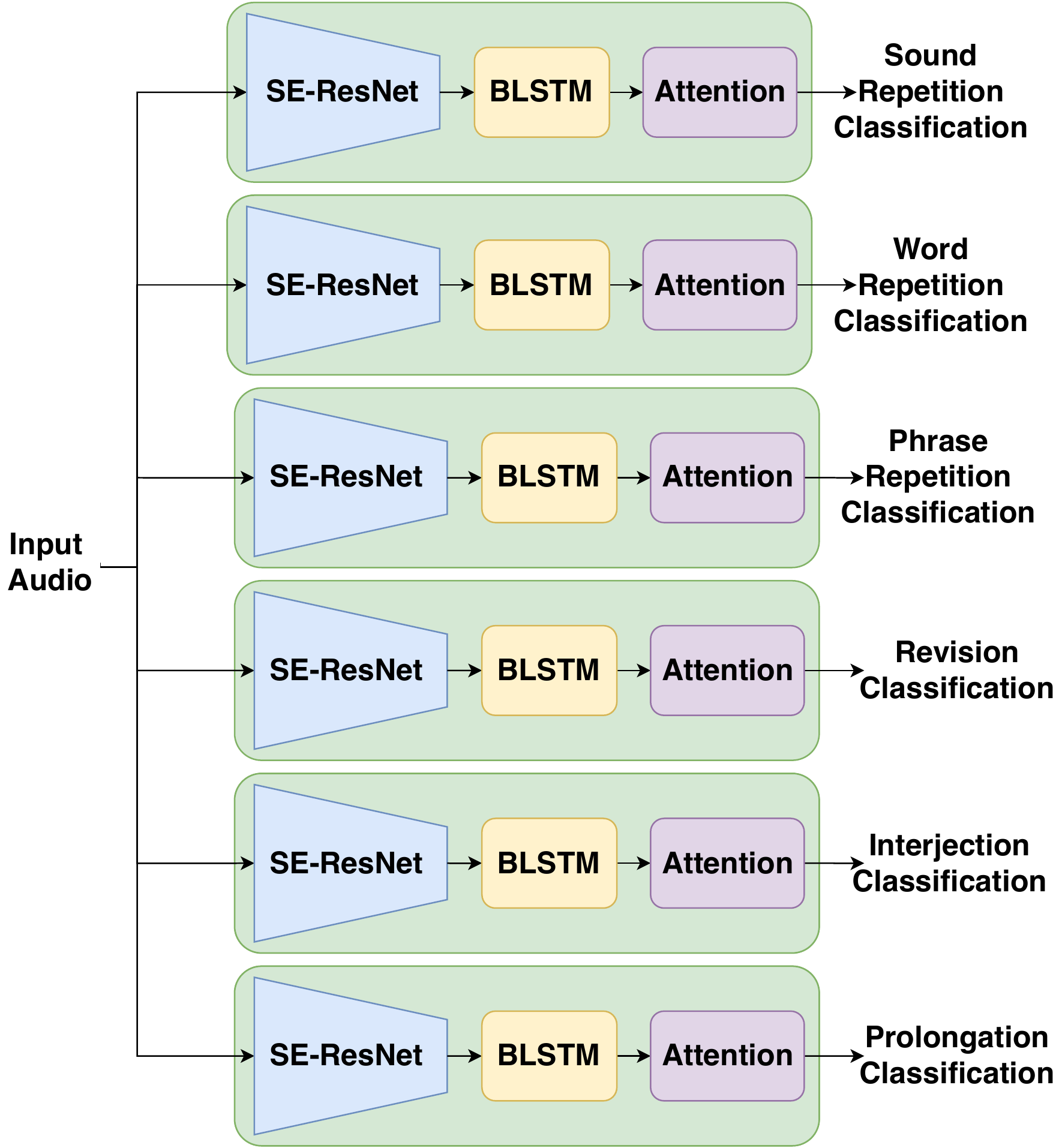}
    \end{center}
\caption{Full model overview using FluentNet for disfluency classification.}
\label{fig:overview}
\end{figure}

Our goal in this section is to design and develop a system that can be used for detecting various types of disfluencies. While one approach to tackle this concept is to design a multi-class problem, another approach is to design an ensemble of single-disfluency detectors. In this paper, given the relatively small size of available stutter datasets, we use the latter approach which can help reduce the complexity of each binary task. 
Accordingly, the goal is to design a single network architecture that can be trained separately to detect different disfluency types with each trained instance, where together they could detect a number of different disfluencies. Figure \ref{fig:overview} shows the overview of our system. The designed network should possess the capability of learning spectral frame-level representations as well as temporal relationships. Moreover, the model should be able to focus on salient parts of the inputs in order to effectively learn the disfluencies and perform accurately. 

\subsection{Proposed Network: FluentNet}
\begin{figure*}[!ht]
    \begin{center}
    \includegraphics[width=0.9\linewidth]{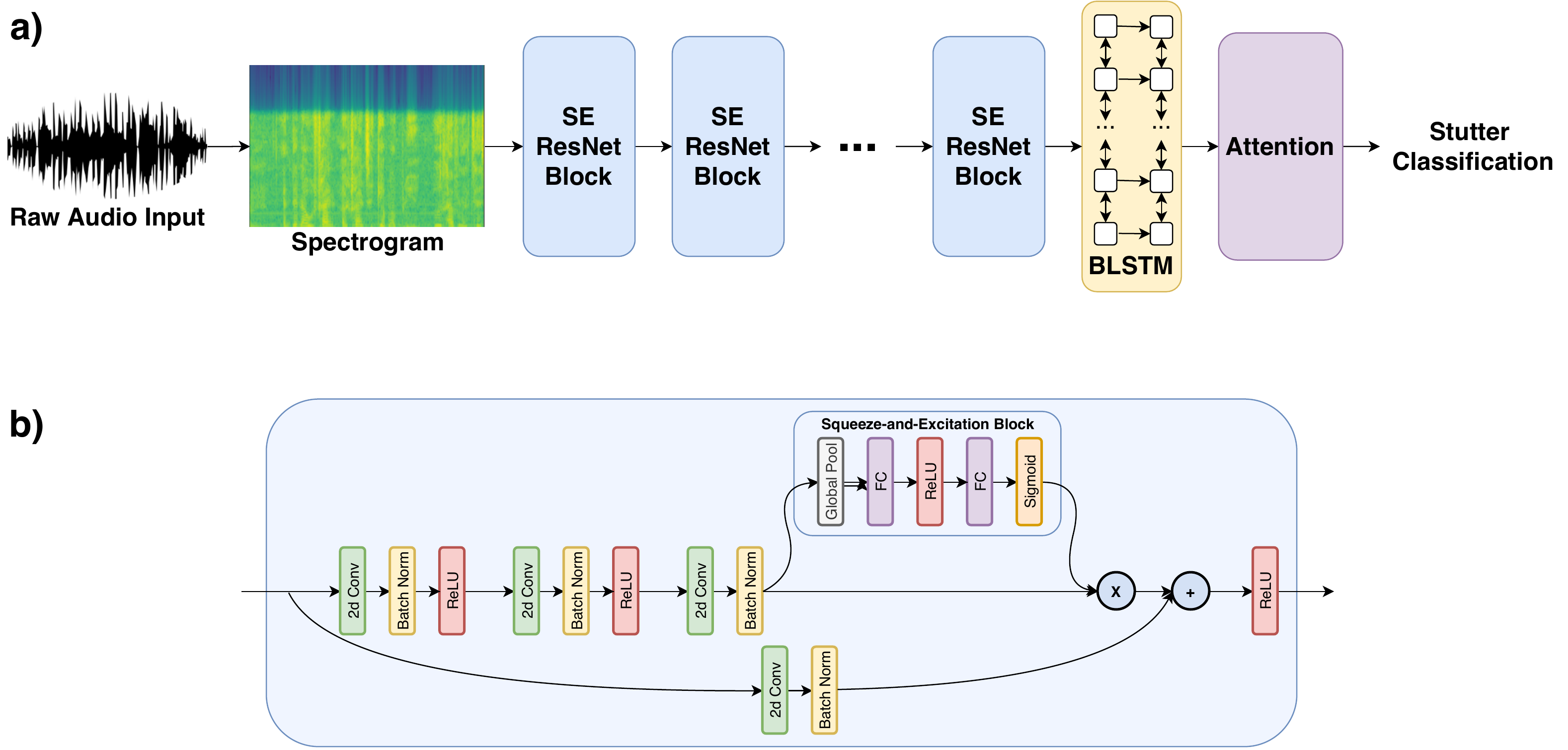}
    \end{center}
\caption{a) A full workflow of FluentNet is presented. This network consists of 8 SE residual blocks, two BLSTM layers, and a global attention mechanism. b) The breakdown of a single SE-ResNet block in FluentNet is presented.}
\label{fig:workflow}
\end{figure*}

We propose an end-to-end network, FluentNet, which uses the short-time Fourier transform (STFT) spectrograms of audio clips as inputs. These inputs are passed through a Squeeze-and-Excitation Residual Network (SE-ResNet) to learn frame-level spectral representations. As most stutter types have distinct spectral and temporal properties, a bidirectional LSTM network is introduced to learn the temporal relationships present among different spectrograms. An attention mechanism is then added to the final recurrent layer to better focus on the necessary features needed for stutter classification. FluentNet's final output reveals a binary classification to detect a specific disfluency type that it has been trained for. The architecture of the network is presented in Figure \ref{fig:workflow}(a). In the following, we describe each of the components of our model in detail.

\subsubsection{Data Representation}

Input audio clips recorded with a sampling rate of 16 \textit{khz} are converted to spectrograms using STFT with 256 filters (frequency bins) to be fed to our end-to-end model. A sample spectrogram can be seen in Figure \ref{fig:workflow} where the colours represent the amplitude of each frequency bin at a given frame, with blue representing lower amplitudes, and green and yellow representing higher amplitudes. Following the common practice in audio-signal processing, a 25 \textit{ms} frame has been used with an overlap of 10 \textit{ms}. 

\subsubsection{Learning Frame-level Spectral Representations}
FluentNet first focuses on learning effective representations from each input spectrogram. To do so, CNN architectures are often used. Though both residual networks \cite{resnet} and squeeze-and-excitation (SE) networks \cite{SENet} are relatively new in the field of deep learning, both have proven to improve on previous state-of-the-art models in a variety of different application areas \cite{szegedy2016inception}, \cite{roy2018}. The ResNet architecture has proven a reliable solution to the vanishing or exploding gradient problems, both common issues when back-propagating through a deep neural network. In many cases, as the model depth increases, the gradients of weights in the model become increasing smaller, or inversely, explosively larger with each layer. This may eventually prevent the gradients from actually changing the weights, or from the weights becoming too large, thus preventing improvements in the model. A ResNet, overcomes this by
utilizing shortcuts all through its CNN blocks
resulting in norm-preserving blocks capable of carrying gradients through very deep models.



Squeeze-and-excitation modules have been recently proposed and have shown to outperform various DNN models using previous CNN architectures, namely VGG and ResNet, as their backbone architectures \cite{SENet}. SE networks were first proposed for image classification, reducing the relative error compared to previous works on the ImageNet dataset by approximately 25\% \cite{SENet}. 

Every kernel within a convolution layer of a CNN results in an added channel (depth) for the output feature map. Whereas recent works have focused on expanding on the spectral relationships within these models \cite{bell2016} \cite{newell2016}, SE-blocks build stronger focus on channel-wise relationships within a CNN. These blocks consist of two major operations. The \textit{squeeze} operation aggregates a feature map across both its height and width resulting in a one-dimensional channel descriptor. The \textit{excitation} operation consists of fully connected layers providing channel-wise weights, which are then applied back to the original feature map.


To exploit the capabilities of both ResNet and SE architectures and learn effective spectral frame-level representations from the input, we use an SE-ResNet model in our end-to-end network. The network consists of 8 SE-ResNet blocks, as shown in Figure \ref{fig:workflow}(a).
Each SE-ResNet block in FluentNet, illustrated in Figure \ref{fig:workflow}(b), consists of three sets of two-dimensional convolution layers, each succeeded by a batch normalization and Rectified Linear Unit (ReLU) activation function. A separate residual connection shares the same input as the block's non-identity branch, and is added back to the non-identity branch before the final ReLU function, but after the SE unit (described below). Each residual connection contains a convolution layer followed by batch normalization. The Squeeze-and-Excitation unit within each SE-ResNet block begins with a global pooling layer. The output is then passed through two fully connected layers: the first followed by a ReLU activation function, and the second succeeded with a sigmoid gating function. The main convolution branch is scaled with the output of the SE unit using channel-wise multiplication.


\subsubsection{Learning Temporal Relationships}
In order to learn the temporal relationships between the representations learned from the input spectrogram, we use an RNN. In particular, LSTM \cite{LSTM} networks have shown to be effective for this purpose in the past and are widely used for learning sequences of spectral representations obtained from consecutive segments of time-series data \cite{ma2018lstm} \cite{kim2018forecasting} \cite{li2020}.

Each LSTM unit contains a cell state, which holds information contained in previous units allowing the network to learn temporal relationships. This cell state is part of the LSTM's memory unit, where there lie several gates that together control which information from inputs, as well as from the previous cell and hidden states, will be used to generate the current cell and hidden states. Namely, the forget gate, $f_t$, and input gate, $i_t$, are utilized to learn what information from each of these respective states will be saved within the new current state, $C_t$. This is shown by the following equations:
\begin{equation}
    f_t = \sigma(W_f \cdot [h_{t-1}, x_t] + b_f)
\end{equation}
\begin{equation}
    i_t = \sigma(W_i \cdot [h_{t-1}, x_t] + b_i)
\end{equation}
\begin{equation}
    C_t = f_t \* \ast C_{t-1} + i_t * tanh(W_C \cdot [h_{t-1}, x_t] + b_C)
\end{equation}
where $\sigma$ represents the sigmoid function, and the $*$ operator denotes point-wise multiplication. This new cell state, along with an output gate, $o_t$, are used to generate the hidden state of the unit, $h_t$, as represented by:
\begin{equation}
    o_t = \sigma(W_o \cdot [h_{t-1}, x_t] + b_o)
\end{equation}
\begin{equation}
    h_t = o_t * tanh(C_t) 
\end{equation}

The cell state and hidden state are then passed to successive LSTM units, allowing the network to learn long-term dependencies.



We used a BLSTM network \cite{BiLSTM} in FluentNet. 
BLSTMs consist of two LSTMs advancing in opposite directions, maximizing the available context from relationships of both the past and future. The outputs of these two networks are multiplied together into a single output layer. FluentNet consists of two consecutive BLSTMs, each utilizing LSTM cells with 512 hidden units. A dropout \cite{dropout} of 20\% was also applied to each of these recurrent layers. To avoid overfitting given the size of the dataset, the randomly masked neurons caused by dropout forces the model to be trained using a sparse representation of the given data.

\subsubsection{Attention} 
\begin{figure}[!t]
    \begin{center}
    \includegraphics[width=0.7\columnwidth]{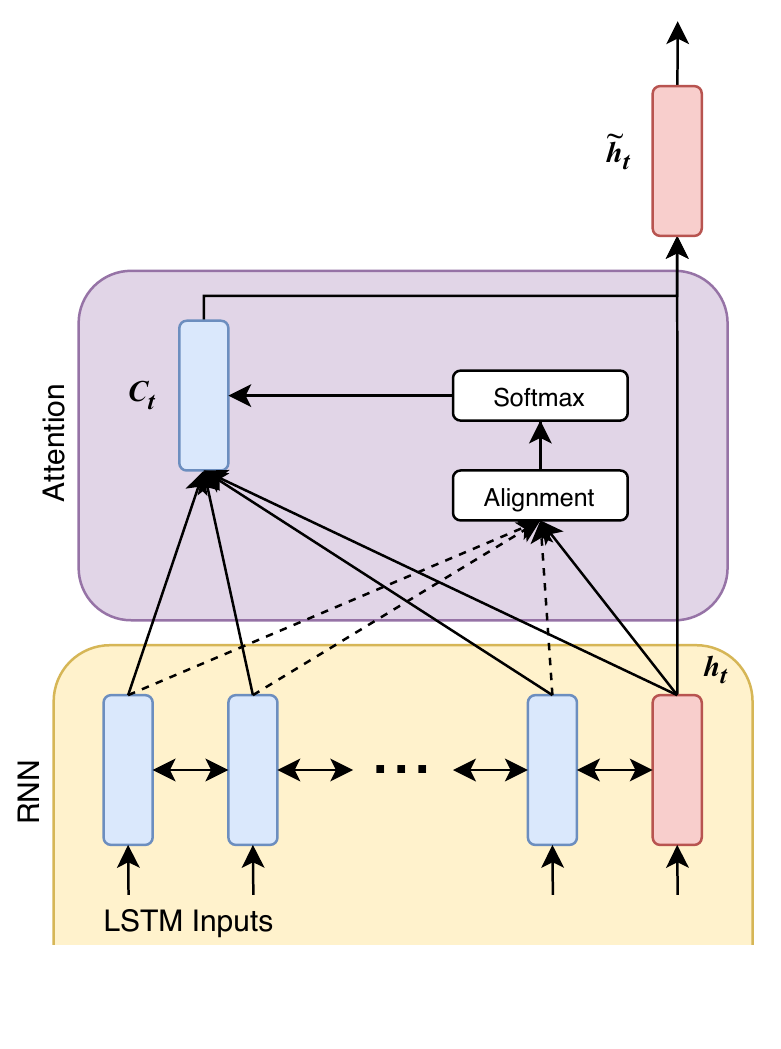}
    \end{center}
\caption{Global attention addition to binary classifier of recurrent network.}
\label{fig:attention}
\end{figure}

The recent introduction of attention mechanisms \cite{bahdanau2014} and its subsequent variations \cite{FEFA2020} have allowed for added focus on more salient sections of the learned embedding. These mechanisms have recently been applied to speech recognition models to better focus on strong emotional characteristics within utterances \cite{mirsamadi2017} \cite{sun2019}, and have similarly been used in FluentNet to improve focus on specific parts of utterances with disfluent attributes. FluentNet uses global attention \cite{luong2015}, which incorporates all hidden state values of the encoder (in this case the BLSTM). A diagram showing the attention model is presented in Figure \ref{fig:attention}. 

The final output value of the second layer of the BLSTM, $h_t$, as well as a context vector, $C_t$, derived through the use of the attention mechanism are used to generate FluentNet's final classification, $\tilde{h}_t$. This is done by applying a tanh activation function, as shown by: 
\begin{equation}
\tilde{h}_t = tanh(W_c[C_t; h_t])
\end{equation}

The context vector of the global attention is the weighted sum of all hidden state outputs of the encoder. An alignment vector, generated as a relation between $h_t$ and each hidden state value is passed through a softmax layer, which is then used to represent the weights to the context vector. Dot product was used as the alignment score function for this attention mechanism. The calculation for the context vector can be represented by:
\begin{equation}
C_t = \sum_{i = 1}^{t} \bar{h}_i (\frac{e^{h_t^\top \cdot \bar{h}_i}}{\sum_{i` = 1}^{t} e^{h_t^\top \cdot \bar{h}_{i`}}})
\end{equation}
where $\bar{h}_i$ represents the $i$th BLSTM hidden state's output.

\subsection{Implementation} \label{Implementation Details}
FluentNet was implemented using Keras \cite{chollet2015keras} with a Tensorflow \cite{abadi2016tensorflow} backend. The model was trained with a learning rate of $10^{-4}$ yielded the strongest results. A root mean square propagation (RMSProp) optimizer, and a binary cross-entropy loss function were used. All experiments were trained using an Nvidia GeForce GTX 1080 Ti GPU. Python's Librosa library \cite{mcfee2015librosa} was used for audio importing and manipulation towards creating our synthetic dataset as described later. Each STFT spectrogram was generated using four-second audio clips. This length of time can encapsulate any stutter apparent in the dataset, with no stutters lasting longer than four seconds.

\section{Experiments}  \label{Experiments}
\subsection{Datasets} 

Despite an abundance of datasets for speech-related tasks such as ASR and speaker recognition \cite{LibriSpeech} \cite{TIMIT} \cite{voxceleb}, there is a clear lack of corpora that are focused on speech disfluencies. An ideal speech disfluency dataset would require the labelling and categorization of each existing disfluent utterance. In this paper, to tackle this problem, in addition to using the UCLASS dataset which is a commonly used stuttered speech corpus \cite{Chee2009_MFCC} \cite{ai2012} \cite{interspeech2018}, a second dataset was created through adding speech disfluencies into clean speech. This synthetic corpus contributes a drastic expansion to the available training and testing data for disfluency classification. Through the following subsections, we describe the UCLASS dataset used in our study, as well as the approach for creating the synthetic dataset, LibriStutter, which we created using the original non-stuttered LibriSpeech dataset.

\subsubsection{UCLASS}
The University College London’s Archive of Stuttered Speech (UCLASS) \cite{UCLASS} is the most commonly used dataset for disfluency-related studies with machine learning. This corpus came in two releases, in 2004 and 2008, from the university's Division of Psychology and Language Sciences. The dataset consists of 457 audio recordings including monologues, readings, and conversations of children with known stutter disfluency issues. Of those recordings, a select few contain written transcriptions of their respective audio files; these were either standard, phonetic or orthographic transcriptions.
Orthographic format is the best option for manual labelling of the dataset for disfluency as they try to transcribe the exact sounds uttered by the speaker in the form of standard alphabet. This helps to identify the presence of disfluency in an utterance more easily. 
The resulting applicable data consisted of 25 unique conversations between an examiner and a child between the ages of 8 and 18, totalling to just over one hour of audio.




In order to pair the utterances with their transcriptions, each audio file and its corresponding orthographic transcription were passed through a forced time alignment tool. The resulting table related each alphabetical token in the transcription to its matching timestamp within the audio. This process was then manually checked for outlaying utterances not matching their transcriptions. 

The provided orthographic transcriptions only flagged the existence of disfluencies (through the use of capitalization), but gave no information towards a disfluency type. To build a more detailed dataset and be able to classify the type of disfluency, all utterances were manually labelled as one of the seven represented classes for our model. These included clean (no stutter), sound repetitions, word repetitions, phrase repetitions, revisions, interjections, and prolongations. The annotation methods applied in \cite{yairiambrose} and \cite{justeandrade} were used as guidelines when manually categorizing each utterance. 
Out of the 8 disfluencies, 6 were used: sound, word, and phrase repetitions, as well as revisions, interjections, and prolongations.. Of the usable audio in the dataset, only three instances of `part-word repetitions' appeared, lacking sufficient positive training samples to feasibly classify these types of stutters. As `block disfluencies' exist as the absence of sound, they could not feasibly be represented in the orthographic transcriptions, which represent how utterances are performed.

\subsubsection{LibriStutter}
The 2015 LibriSpeech ASR corpus by Panayotov et al. \cite{LibriSpeech} includes 1000 hours of prompted English speech extracted from audio books derived from the LibriVox project. We used this dataset as the basis for our synthetic stutter dataset, which we name LibriStutter. LibriStutter's creation compensates for two shortcomings of the UCLASS corpus: the small amount of labelled stuttered speech data available and the imbalance of the dataset (several disfluency types in UCLASS consisted of a small number of samples). 

To allow for a manageable size for LibriStutter and feasible training times, we used a subset of LibriSpeech and set the size of LibriStutter to 20 hours. LibriStutter includes synthetic stutters for sound repetitions, word repetitions, phrase repetitions, prolongations, and interjections. We generated these stutter types by sampling the audio within the same utterance, the details of which are described below. Revisions were excluded from LibriStutter, as this disfluency type requires the speaker to change and revise what was initially said. This would require added speech through the use of complex language models and voice augmentation tools to mimic the revised phrase, both of which fall out of scope for this project.


For each audio file selected from the LibriSpeech dataset, we used the Google Cloud Speech-to-Text API \cite{googleTTS} to generate a timestamp corresponding to each spoken word. For every four-second window of speech within a given audio file, either a random disfluency type was inserted and labelled accordingly, or alternatively left clean. Each disfluency type underwent a number of processes to best simulate natural stutters. 

All repetition stutters relied upon copying existing audio segments already present within each audio file. Sound repetitions were generated by copying the first fraction of a random spoken word within the sample and repeating this short utterance a several times before said word. Although repetitions of sounds can occur at the end of words, known as word-final disfluencies, this is rarely the case \cite{van2005}. One to three repeated sound utterances were added in each stuttered word. After each instance of the repeated sound, a random empty pause duration of 100 to 350 \textit{ms} was appended as this range sounded most natural. Inserted audio may leave sharp cutoffs, especially part-way through an utterances. To avoid this, interpolation was used to smooth the added audio's transition into the existing clip. 

Both word and phrase repetitions underwent similar processes to that of sound repetitions. For word repetitions we repeated one to two copies of a randomly selected word before the original utterance. For phrase repetitions, a similar approach was taken, where instead of repeating a particular word, a phrase consisting of two to three words were repeated. The same pause duration and interpolation techniques used for sound repetitions were applied to both word and phrase repetition disfluencies.

Prolongations consist of sustained sounds, primarily at the end of a word. To mimic this behaviour, the last portion of a word was stretched to simulate prolonged speech. For a randomly chosen word, the latter 20\% of the signal was stretched by a factor of 5. This prolonged speech segment replaced the original word ending. As applying time stretching to audio results in a drop in pitch, pitch shifting was used to realign the pitch with the original audio. The average pitch of the given speech segment was used to normalize the disfluent utterance.

Unlike the aforementioned classes, interjection disfluencies cannot be created from existing speech within a sample as it requires the addition of filler words absent from the original audio (for example `umm'). Multiple samples of common filler words from the UCLASS were isolated and saved separately to create a pool of interjections. To create interjection disfluencies, a random filler word from this pool was inserted between two random words, followed by a short empty pause. The same pitch scaling and normalization method as used for prolongations was applies to match the pitches between the interjection and audio clip. Interpolation was used as in repetition disfluencies to smooth sharp cutoffs caused by the added utterance.

\begin{figure}[!ht]
    \begin{center}
    \includegraphics[width=1\linewidth]{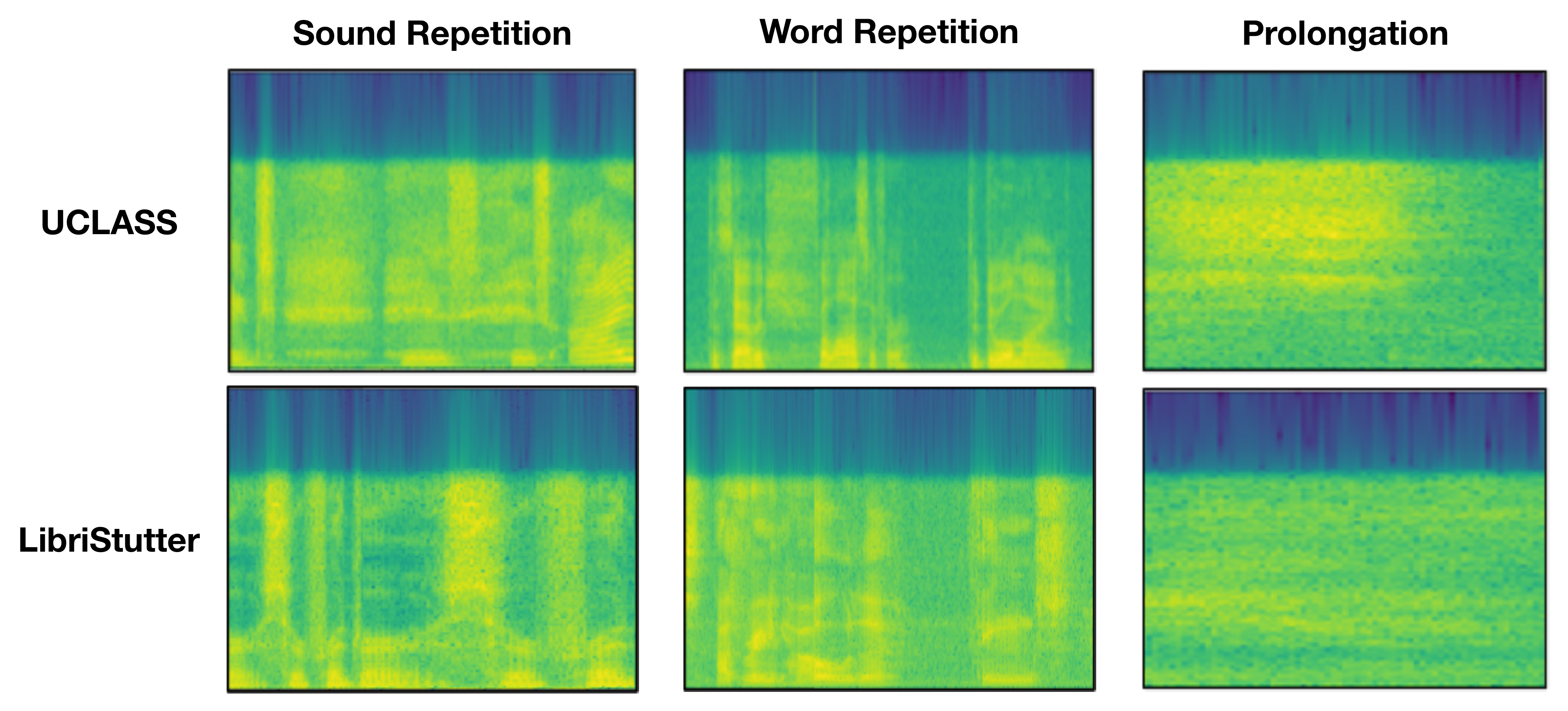}
    \end{center}
\caption{Spectrograms of the same stutters found in the UCLASS dataset and generated in the LibriStutter dataset.}
\label{fig:spectrogram_images}
\end{figure}

\begin{table}
\begin{center}
\footnotesize
\caption{Cosine similarity between a UCLASS dataset stutter and a matching LibriStutter stutter, as well as the average of 100 random samples from the LibriStutter dataset.}
\label{table: cosine_similarity}
\begin{tabular}{ l c c }
\hline
 Stutter & UCLASS vs. LibriStutter & UCLASS vs. Random\\ \hline\hline
Sound Repetition & $3.73893\mathrm{e}{-3}$ & $2.58116\mathrm{e}{-4}$\\
Word Repetition & $3.14077\mathrm{e}{-3}$ & $2.61084\mathrm{e}{-4}$\\
Prolongation & $7.70236\mathrm{e}{-3}$  & $2.57234\mathrm{e}{-4}$\\
\hline
\end{tabular}%
\end{center}
\end{table}


To ensure that sufficient realism was incorporated into the dataset, a registered speech language pathologist 
was consulted for this project. Nonetheless, it should be mentioned that despite our attention to creating a perceptually valid and realistic dataset, the notion of ``realism'' itself is not a focus of this dataset. Instead, much like synthetic datasets in other areas such as image processing, the aim is for the dataset to be \textit{valid enough} such that machine learning and deep learning methods can be trained and evaluated with, and later on transferred to real large-scale datasets [in the future] with little to no adjustments to the model architectures. 




Figure \ref{fig:spectrogram_images} displays side by side comparisons of spectrograms of real stuttered data from the UCLASS dataset, and artificial stutters from LibriStutter. Each pairing represents a single stutter type, with the same word or sound being spoken in each. It can be observed that the UCLASS stutter samples and their corresponding LibriStutter examples show clear similarities.
Moreover, to numerically compare the samples, cosine similarity \cite{cosine_similarity} was calculated between the UCLASS and LibriStutter spectrogram samples shown earlier.
To add relevance to these values, a second comparison was performed for each UCLASS spectrogram with respect to 100 random samples from the LibriStutter dataset, and the average score was used as the represented comparison value. These scores are summarized in Table \ref{table: cosine_similarity}. We observe that the UCLASS cosine similarity scores corresponding to the matching LibriStutter samples are noticeably (approximately between 10$\times$ to 30$\times$) greater than those compared to random audio samples, confirming that the disfluent utterances contained in LibriStutter share phonetic similarities with real stuttered samples, empirically showing the similarity between a few sample real and synthesized stutters.

The LibriStutter dataset consists of approximately 20 hours of speech data
from the LibriSpeech train-clean-100 (training set of 100 hours ``clean'' speech). In turn, LibriStutter shares a similar make up to that of its predecessor. It consists of \textit{disfluent} prompted English speech from audiobooks. 
It also contains 23 male and 27 female speakers, with an approximate 53\% of the audio coming from males, and 47\% from females. There are 15000 disfluencies in this dataset, with equal counts for each of the five disfluency types: 3000 sound, word, and phrase repetitions, as well as prolongations and interjections. Each audio file has a corresponding CSV file containing each word or utterance spoken, the start and end time of the utterance, and its disfluency type, if any. 



\subsection{Benchmarks}
For a thorough analysis of our results, we compare the results obtained by the proposed FluentNet to a number of other models. In particular, we employ two type of solutions for comparison purposes. First, we compare our results to related works and the state-of-the-art as follows:

    \textbf{Alharbi et al. \cite{interspeech2018}:} This work conducted classification of sound repetitions, word repetitions, revisions, and prolongations on the UCLASS dataset through the application of two different methods. First, an original speech prompt was aligned, and then passed to a task-oriented FST to generate word lattices. These lattices were used to detect repeated part-words, words, and phrases within the sample. 
    This method scored perfect results on word repetition classification,
    though the results on sound repetitions and revisions proved much weaker. To classify prolongation stutters, an autocorrelation algorithm consisting of two thresholds was used: the first to detect speech with similar amplitudes (sustained speech), and another dynamic threshold to decide whether the duration of similar speech would be considered a prolongation. Using this algorithm, perfect prolongation classification was achieved.
    
    \textbf{Chen et al. \cite{Chen2020}:} A CT-Transformer was designed to conduct repetition and interjection disfluency detection on an in-house Chinese speech dataset. Both word and position embeddings of a provided audio sample were passed through a series of CT self attention layers and fully connected layers. This work was able to achieve an overall disfluency classification miss rate of 38.5\% (F1 score of 70.5). Notably, this is one of the few works to have attempted interjection disfluency classification, yielding a miss rate of 25.1\%. Note that the performance on repetition disfluencies encompasses all repetition-type stutters, including sound, word, and phrase repetitions, as well as revisions.
    
    \textbf{Kourkounakis et al. \cite{kourkounakis2020}:} As opposed to other current models focusing on ASR and language models, our previous work proposed a model relying solely on acoustic and phonetic features, allowing for the classification of several multiple disfluencies types without the need for speech recognition methods. This model applied a deep residual network, consisting of 6 residual blocks (18 convolution layers) and two bidirectional long short-term memory layers to classify six different types of stutters. This work achieved an average miss rate of 10.03\% on the UCLASS dataset, and sustained strong accuracy and miss rates across all stutter types, prominently word repetitions and revisions.

    \textbf{Zayats et al. \cite{zayats2016}:} A recurrent network was used to classify repetition disfluencies within the Switchboard corpus. It consists of a BLSTM followed by an ILP post processing method. The input embedding to this network consisted of a vector containing each word's index, part of speech, as well as 18 other disfluency-based features. The method achieved a miss rate of 19.4\% across all repetitions disfluencies.

    \textbf{Villegas et al. \cite{villegas2019}:} This model was used a reference to compare the effectiveness of repository signals towards stutter classification. These features included the means, standard deviations, and distances of respiratory volume, respiratory flow, and heart rate. Sixty-eight participants were used to generate the data for their experiments. The best performing model in this work was an MLP with 40 hidden layers, resulting in a 82.6\% average classification accuracy between block and non-block type stutters.

    \textbf{Dash et al. \cite{dash2018}:} This method passed the maximum amplitude of the provided audio sample through a neural network to generate a custom threshold for each sample, trained on a set of 60 speech samples. This amplitude threshold was used to remove any perceived prolongations and interjections. The audio was then passed the audio through a SST tool, which allowed for the removal of any repeated words, phrases, or characters, achieving an overall stutter classification of 86\% on a test set of 50 speech segments.
    

Note that the latter three works only provide results on a group of disfluency types \cite{zayats2016}, a single disfluency type \cite{villegas2019}, or overall stutter classification \cite{dash2018}. As such, only their average disfluency classification results could be compared. Moreover, these works (\cite{Chen2020}, \cite{zayats2016}, \cite{villegas2019}, and \cite{dash2018}) have not used the UCLASS dataset, therefore the comparisons should be taken cautiously.

Next, we also compare the performance of our solution to a number of other models for benchmarking purposes. These models were selected due to their popularity for time-series learning and their hyperparameters of these models are all tuned to obtain the best possible results given their architectures. These benchmarks are as follows:
    (\textit{i}) VGG-16 (Benchmark 1): VGG-16 \cite{vgg} 
    consists of 16 convolutional or fully connected layers, comprised of groups of two or three convolution layers with ReLU activation, with each grouping being followed by a max pooling layer.
    The model concludes with three fully connected layers and a final softmax function. 
    (\textit{ii}) VGG-19 (Benchmark 2): 
    This network is very similar to its VGG-16 counterpart, with the only difference being an addition of three more convolution layers spread throughout the model. 
    (\textit{iii}) ResNet-18 (Benchmark 3): 
    ResNet-18 was chosen as a benchmark, which contains 18 layers: eight consecutive residual blocks each containing two convolutional layers with ReLU activation, followed by an average pooling layer and a final fully connected layer. 

\section{Results and Analysis} \label{Results}

\subsection{Validation}
In order to rigorously test FluentNet on the UCLASS dataset, a leave-one-subject-out (LOSO) cross validation method was used. The results of models tested on this dataset are represented as the average between 25 experiments, each consisting of audio samples from 24 of the participants as training data, and a unique single participant's audio as a test set. A 10-fold cross validation method was used on the LibriStutter dataset with a random 90\% subset of the samples from each stutter being used for training along with 90\% of the clean samples chosen randomly. The remaining 10\% of both clean and stuttered samples were used for testing. All experiments were trained over 30 epochs, with minimal change in loss seen in further epochs.



The two metrics used to measure the performance of the aforementioned experiments were miss rate and accuracy. Miss rate (1 - \textit{recall}) is used to determine the proportion of disfluencies which were incorrectly classified by the model. To balance out any bias this metric may hold, accuracy was used as a second performance metric. 

\begin{table*}
\begin{center}
\footnotesize
\caption {Percent miss rates (MR) and accuracy (Acc) of the six stutter types trained on the UCLASS dataset.}
\label{table: results_uclass}
\resizebox{\linewidth}{!}{\begin{tabular}{lll|cc|cc|cc|cc|cc|cc}
\hline
 & & & \multicolumn{2}{c|}{S} & \multicolumn{2}{c|}{W} & \multicolumn{2}{c|}{PH} & \multicolumn{2}{c|}{I} & \multicolumn{2}{c|}{PR} & \multicolumn{2}{c}{R} \\
Paper & Method & Dataset & MR$\downarrow$ & Acc.$\uparrow$ & MR$\downarrow$ & Acc.$\uparrow$ & MR$\downarrow$ & Acc.$\uparrow$ & MR$\downarrow$ & Acc.$\uparrow$ & MR$\downarrow$ & Acc.$\uparrow$ & MR$\downarrow$ & Acc.$\uparrow$\\ \hline\hline
Alharbi et al. \cite{interspeech2018} & Word Lattice & UCLASS & 60 & -- & \textbf{0} & -- & \cellcolor{gray!25} & \cellcolor{gray!25} & \cellcolor{gray!25} & \cellcolor{gray!25} & \textbf{0} & --  & 25 & --\\
Kourkounakis et al. \cite{kourkounakis2020} & ResNet+BLSTM & UCLASS & 18.10 & 84.10 & 3.20 & \textbf{96.60} & 4.46 & 95.54 & 25.12 & 81.40 & 5.92 & 94.08 & 2.86 & 97.14 \\
Benchmark 1 & VGG-16 & UCLASS & 20.80 & 81.03 & 6.54 & 93.01 & 12.82 & 87.91 & 28.44 & 72.03 & 9.04 & 90.83 & 5.20 & 94.90\\
Benchmark 2 & VGG-19 & UCLASS & 19.41 & 81.35 & 5.22 & 95.42 & 10.13 & 91.60 & 26.06 & 73.64 & 5.72 & 94.21 & 4.72 & 96.32 \\
Benchmark 3 & ResNet-18 & UCLASS & 19.51 & 81.38 & 5.26 & 94.50 & 7.32 & 94.01 & 25.55 & 76.38 & 7.02 & 93.22 & 5.16 & 94.74\\
\textbf{Ours} & \textbf{FluentNet} & UCLASS & \textbf{16.78} & \textbf{84.46} & 3.43 & 96.57 & \textbf{3.86} & \textbf{96.26} & \textbf{24.05} & \textbf{81.95} & 5.34 & \textbf{94.89} & \textbf{2.62} & \textbf{97.38} \\
\hline
Kourkounakis et al. \cite{kourkounakis2020} & ResNet+BLSTM & LibriStutter & 19.23 & 79.80 & 5.17 & 92.52 & 6.12 & 92.52 & 31.49 & 69.22 & 9.80 & 89.44 & \cellcolor{gray!25} & \cellcolor{gray!25}\\
Benchmark 1 & VGG-16 & LibriStutter & 20.97 & 79.33 & 6.27 & 92.74 & 8.90 & 91.94 & 36.47 & 64.05 & 10.63 & 89.10 & \cellcolor{gray!25} & \cellcolor{gray!25}\\
Benchmark 2 & VGG-19 & LibriStutter & 20.79 & 79.66 & 6.45 & 93.44 & 7.92 & 92.44 & 34.46 & 66.92 & 10.78 & 89.98 & \cellcolor{gray!25} & \cellcolor{gray!25}\\
Benchmark 3 & ResNet-18 & LibriStutter & 22.47 & 78.71 & 6.22 & 92.70 & 6.74 & 93.36 & 35.56 & 64.78 & 10.52 & 90.32 & \cellcolor{gray!25} & \cellcolor{gray!25}\\
 \textbf{Ours} & \textbf{FluentNet} & LibriStutter & \textbf{17.65} & \textbf{82.24} & \textbf{4.11} & \textbf{94.69} & \textbf{5.71} & \textbf{94.32} & \textbf{29.78} & \textbf{70.12} & \textbf{7.88} & \textbf{92.14}& \cellcolor{gray!25} & \cellcolor{gray!25}\\
\hline
\end{tabular}}
\end{center}
\end{table*}

\subsection{Performance and Comparison}
The results of our model for recognition of each stutter type are presented for the UCLASS and LibriStutter datasets in Table \ref{table: results_uclass}. FluentNet achieves strong results against all the disfluency types within both datasets, outperforming nearly all of the related work as well as the benchmark models. 

As some previous works have been designed to tackle specific disfluency types as opposed to a general solution for detecting different types of disfluencies, a few of FluentNet's individual class accuracies do not surpass previous works', namely word repetitions and prolongation. In particular, the work by Alharbi et al. \cite{interspeech2018} offers perfect word repetition classification, as word lattices can easily identify two words repeated one after the other. Amplitude thresholding also proves to be a successful prolongation classification method. It should be noted that FluentNet does achieve strong results for these disfluency types as well.
Notably, our work is one of the only ones that has attempted classification of interjection disfluencies. 
These disfluent utterances lack the unique phonetic and temporal patterns that, for instance, repetition or prolongation disfluencies contain. Moreover, they may be present as a combination of other disfluency types, for example an interjection can be both prolonged or repeated. For these reasons, interjections remain the hardest category, with a 24.05\% and 29.78\% miss rate on the UCLASS an LibriStutter datasets, respectively. Nonetheless, FluentNet provides good results, especially given that interjections have been historically avoided.

The task-oriented lattices generated in \cite{interspeech2018} show strong performance on word repetitions and prolongations, but struggle to detect sound repetitions and revision. Likewise, as is presented in \cite{Chen2020}, the CT-Transformer yields a comparable interjection classification miss rate to that of FluentNet. However, when the same model is applied to repetition stutters, the performance of the model drops severely, hindering its overall disfluency detection capabilities. The use of an attention-based transformer proves a viable method of classifying interjection disfluencies, however as the results suggest, the convolutional and recurrent architecture in FluentNet allows for effective representations to be learned for interjection disfluencies alongside repetitions and prolongations.


FluentNet's achievements surpass our previous work's across all disfluency types on the Libristutter dataset, and all but word repetition accuracy on the UCLASS dataset. The results show a greater margin of improvement against the LibriStutter dataset as compared to UCLASS between the two models. Notably, word repetitions and prolongation relay a decrease in miss rate of approximately 20\% between FluentNet and \cite{kourkounakis2020}. This implies the SE and attention mechanisms assist in better representing the disfluent utterances within stuttered speech found in the synthetic dataset.

An interesting observation is that LibriStutter proves a more difficult dataset compared to UCLASS as evident by the lower performance of all the solutions including FluentNet. This is likely due to the fact that given the large number of controllable parameters for each stutter type, LibriStutter is likely to contain a larger variance of stutter styles and variations, resulting in a more difficult problem space.


Table \ref{table: avg_uclass} presents the overall performance of our model with respect to all disfluency types on UCLASS and LibriStutter datasets. The results are compared with other works on respective datasets, and the benchmarks which we implemented for comparison purposes.
We observe that FluentNet achieves average miss rates and accuracy of 9.35\% and 91.75\%on the UCLASS dataset, surpassing the other models and \textit{setting a new state-of-the-art}. A similar trend can be seen for the LibriStutter dataset where FluentNet outperforms the previous model along with all the benchmark models. 


The BLSTM used in \cite{zayats2016} yields successful results towards repetition stutter classification by learning temporal relationships between words, however it remains impaired by its reliance solely on lexical model inputs. 
On the other hand, as shown by the results, FluentNet is better able to learn these phonetic details through the spectral and temporal representations of speech.


The work from \cite{dash2018} uses similar classification techniques to \cite{interspeech2018}, however improves upon the thresholding technique with the addition of a neural neural network. Though achieving an average accuracy of 86\% across the same disfluency types used in this work, FluentNet remains a stronger model given its effective spectral frame-level and temporal embeddings. Nonetheless, the results of this work contains only a single overall accuracy value across all of repetition, interjection, and prolongation disfluency detection. Little is discussed on the origin and makeup of the dataset used.


Of the benchmark models without an RNN component, ResNet performs better than both VGG networks for both datasets, indicating that ResNet-style architectures are able to learn effective spectral representations of speech. This further justifies the use of a ResNet as the backbone of our model. Moreover, the addition of the LSTM component to the benchmarks shows that learning the temporal relationships through an RNN contributes to the performance.

\begin{table}
\begin{center}
\caption{Average percent miss rates (MR) and accuracy (Acc) of disfluency classification models.}
\label{table: avg_uclass}
\resizebox{\columnwidth}{!}{\begin{tabular}{l l c c} 
\hline
Paper & Dataset & Ave. MR$\downarrow$ & Ave. Acc.$\uparrow$ \\
\hline\hline
Zayats et al. \cite{zayats2016} & Switchboard & 19.4 & -- \\
Villegas et al. \cite{villegas2019} & Custom & -- & 82.6 \\
Dash et al. \cite{dash2018}  & Custom &  -- & 86 \\
Chen et al. \cite{Chen2020} & Custom & 38.5 & -- \\
\hline
Alharbi et al. \cite{interspeech2018}  & UCLASS & 37 & -- \\
Kourkounakis et al. \cite{kourkounakis2020}  & UCLASS & 10.03 & 91.15\\
Benchmark 1 (VGG-16) & UCLASS & 13.81 & 86.62\\
Benchmark 2 (VGG-19) & UCLASS & 12.21 & 87.92\\
Benchmark 3 (ResNet-18) & UCLASS & 12.14 & 89.14\\
\textbf{FluentNet} & UCLASS & \textbf{9.35} & \textbf{91.75} \\
\hline
Kourkounakis et al. \cite{kourkounakis2020} & LibriStutter & 14.36 & 85.30\\
Benchmark 1 (VGG-16) & LibriStutter & 16.65 & 83.43\\
Benchmark 2 (VGG-19) & LibriStutter & 16.08 & 84.49\\
Benchmark 3 (ResNet-18) & LibriStutter & 16.30 & 83.97\\
\textbf{FluentNet} & LibriStutter & \textbf{13.03} & \textbf{86.70} \\
\hline
\end{tabular}}
\end{center}
\end{table}



\begin{figure}[t!]
    \centering
    \begin{subfigure}[t]{0.5\columnwidth}
        \centering
        \includegraphics[width=1\columnwidth]{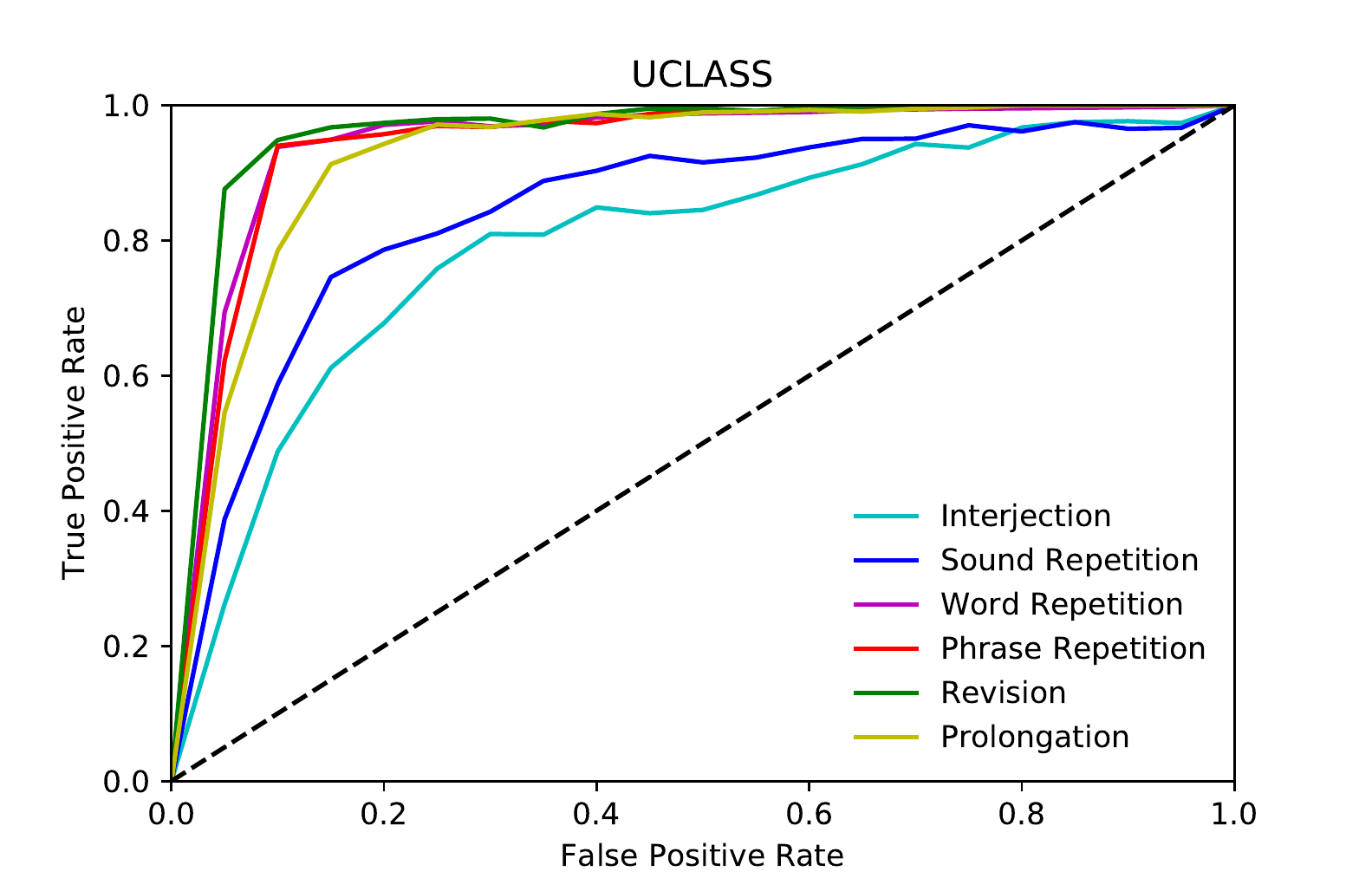}
        \caption{UCLASS}
    \end{subfigure}%
    ~ 
    \begin{subfigure}[t]{0.5\columnwidth}
        \centering
        \includegraphics[width=1\columnwidth]{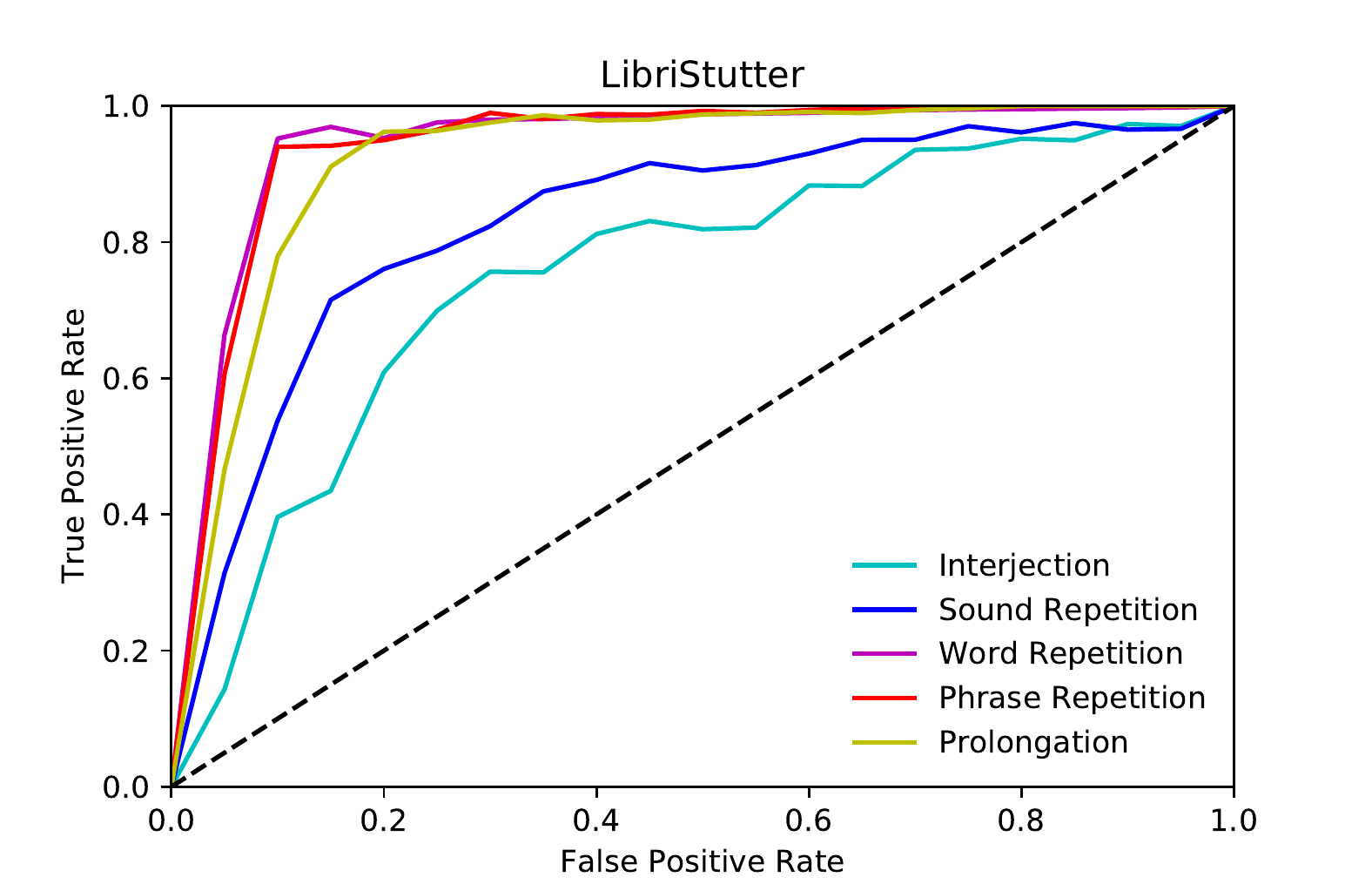}
        \caption{LibriStutter}
    \end{subfigure}
    \caption{ROC curves for each stutter type tested on the UCLASS and LibriStutter datasets.}
\label{fig:roc}
\end{figure}

To further demonstrate the performance of FluentNet, the Receiver Operator Characteristic (ROC) curves were generated for each disfluency class on the UCLASS and LibriStutter datasets, as shown in Figures \ref{fig:roc}(a) and \ref{fig:roc}(b), respectively.
It can be seen that word repetitions, phrase repetitions, revisions, and prolongations reveal very strong classification on both datasets. Both sound repetitions and interjections classification fair weakest, with the LibriStutter dataset, proving to be a more difficult dataset for FluentNet, as previously observed and discussed.

\begin{table*}
\begin{center}
\footnotesize
\caption{Ablation experiment results, providing accuracy (Acc) and miss rates (MR) for each stutter type and model on the UCLASS dataset.}
\label{table: ablation_uclass}
\resizebox{\linewidth}{!}{\begin{tabular}{ll|cc|cc|cc|cc|cc|cc|cc}
\hline
 & & \multicolumn{2}{c|}{S} & \multicolumn{2}{c|}{W} & \multicolumn{2}{c|}{PH} & \multicolumn{2}{c|}{I} & \multicolumn{2}{c|}{PR} & \multicolumn{2}{c|}{R} & \multicolumn{2}{c}{Average}\\
Method & Dataset & MR$\downarrow$ & Acc.$\uparrow$ & MR$\downarrow$ & Acc.$\uparrow$ & MR$\downarrow$ & Acc.$\uparrow$ & MR$\downarrow$ & Acc.$\uparrow$ & MR$\downarrow$ & Acc.$\uparrow$ & MR$\downarrow$ & Acc.$\uparrow$ & MR$\downarrow$ & Acc.$\uparrow$\\ \hline\hline
\textbf{FluentNet} & UCLASS & 16.78 & 84.46 & 3.43 & 96.57 & 3.86 & 96.26 & 24.05 & 81.95 & 5.34 & 94.89 & 2.62 & 97.38 & 9.35 & 91.75\\
w/o Attention & UCLASS & 16.97 & 83.13 & 3.51 & 96.29 & 4.23 & 95.78 & 24.22 & 80.78 & 6.88 & 92.50 & 3.25 & 96.55 & 9.84 & 90.84\\
w/o Squeeze-and-Excitation & UCLASS & 17.37 & 82.01 & 4.82 & 95.34 & 4.81 & 95.17 & 24.59 & 79.84 & 6.22 & 93.10 & 3.14 & 96.98 & 10.16 & 90.41\\
w/o Squeeze-and-Excitation \& Attention & UCLASS & 18.18 & 82.83 & 4.96 & 95.04 & 5.32 & 93.68 & 28.89 & 71.01 & 8.30 & 91.72 & 3.30 & 96.70 & 11.49 & 88.50\\
\hline
\textbf{FluentNet} & LibriStutter & 17.65 & 82.24 & 4.11 & 94.69 & 5.71 & 94.32 & 29.78 & 70.12 & 7.88 & 92.14 & \cellcolor{gray!25} & \cellcolor{gray!25} & 13.03 & 86.70\\
w/o Attention &  LibriStutter & 18.91 & 81.14 & 4.17 & 94.01 & 5.92 & 93.73 & 31.26 & 68.91 & 8.53 & 91.24 &  \cellcolor{gray!25} & \cellcolor{gray!25} & 13.76 & 85.81 \\
w/o Squeeze-and-Excitation &  LibriStutter & 19.11 & 80.72 & 4.95 & 94.60 & 5.87 & 94.15 & 31.14 & 70.02 & 8.80 & 91.28 & \cellcolor{gray!25} & \cellcolor{gray!25} & 13.97 & 86.15\\
w/o Squeeze-and-Excitation \& Attention &  LibriStutter & 19.23 & 79.80 & 5.17 & 92.52 & 6.12 & 92.52 & 31.49 & 69.22 & 9.80 & 89.44 & \cellcolor{gray!25} & \cellcolor{gray!25} & 14.36 & 85.30\\
\hline
\end{tabular}}
\end{center}
\end{table*}

\subsection{Parameters}
Multiple parameters have been tuned in order to maximize the accuracy of FluentNet and the baseline experiments on both datasets. These include convolution window sizes, epochs, and learning rates, among others. Each has been individually tested in order to find the optimal values for the given model. Note that all of FluentNet's hyper-parameters remain the same across all disfluency types. 

Thorough experiments were performed to obtain the optimum architecture of FluentNet. For the SE-ResNet component, we tested a different count of convolution blocks, ranging between 3 to 12, with each block consisting of 3 convolutional layers. Eight blocks were found to be the approximate optimal depth for training the model on the UCLASS dataset. Similarly, we experimented with the use of different number of BLSTM layers, ranging between 0 to 3 layers. The use of 2 layers yielded the best results. Moreover, the use of bi-directional layers proved slightly more effective than uni-directional layers. Lastly, we experimented with a number of different values and strategies for the learning rate where $10^{-4}$ showed the best results.

\begin{figure}[t!]
    \centering
    \begin{subfigure}[t]{0.5\columnwidth}
        \centering
        \includegraphics[width=1\columnwidth]{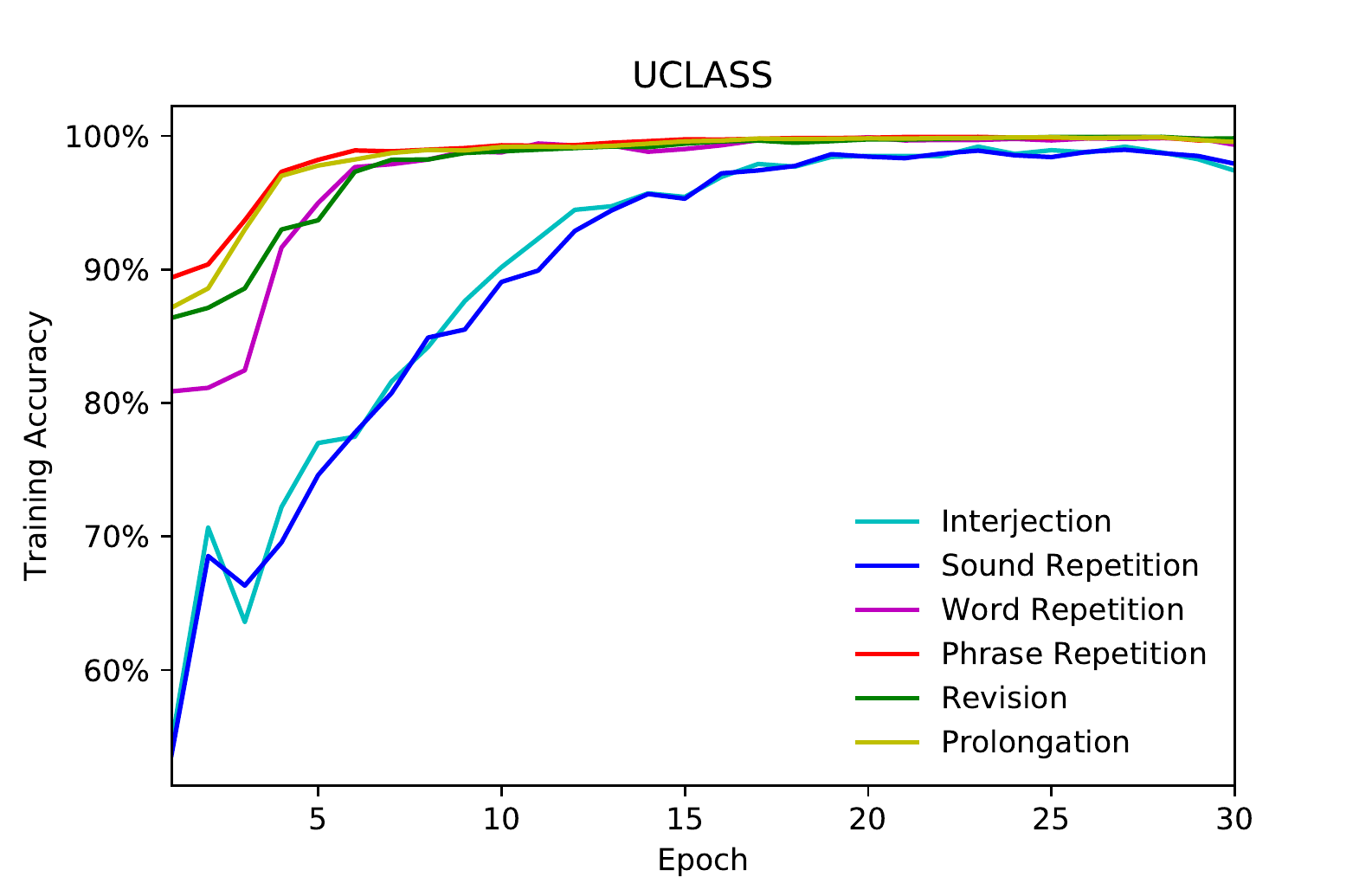}
        \caption{UCLASS}
    \end{subfigure}%
    ~ 
    \begin{subfigure}[t]{0.5\columnwidth}
        \centering
        \includegraphics[width=1\columnwidth]{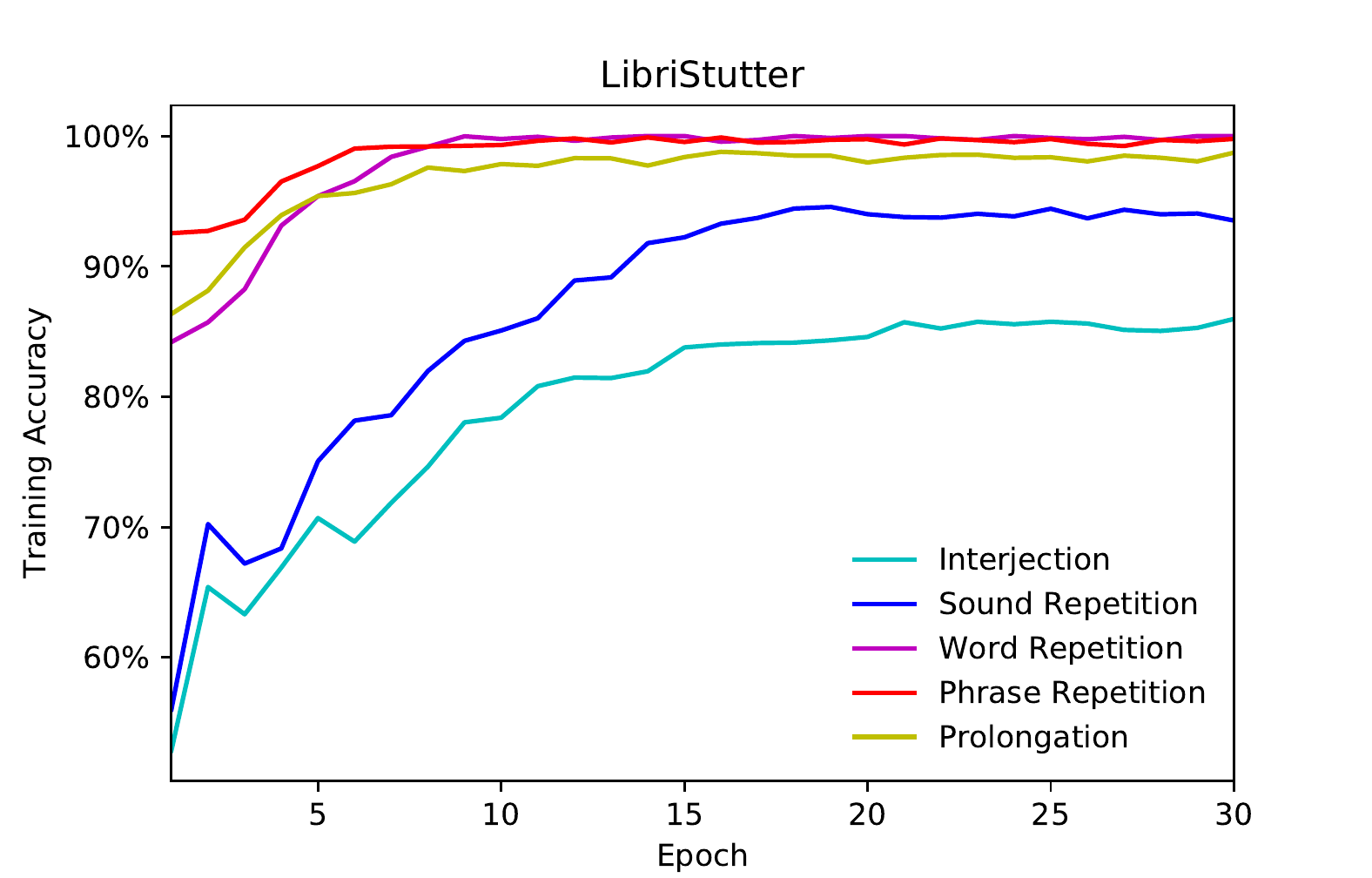}
        \caption{LibriStutter}
    \end{subfigure}
    \caption{Average training accuracy for FluentNet on the considered stuttered types for the UCLASS and LibriStutter datasets.}
\label{fig:training}
\end{figure}

Figures \ref{fig:training}(a) and \ref{fig:training}(b) show FluentNet's performance for each stutter type against different epochs on the UCLASS and LibriStutter datasets, respectively. It can be seen that the training accuracy stabilizes after around 20 epochs. Whereas all disfluencies types in the UCLASS dataset approach perfect training accuracy, training accuracy plateaus at much lower accuracies for interjections and sound repetitions within the LibriStutter dataset.

\subsection{Ablation Experiments}
To further analyze FluentNet, an ablation study was done in order to systematically evaluate how each component contributes towards the overall performance. Both the SE portion and attention mechanisms were removed, individually and together, in order to analyse the relationship between their absences, and how these affect both accuracy and miss rates for each disfluency class. The ablation results for both the UCLASS and LibriStutter datasets can be seen summarized in Table \ref{table: ablation_uclass}. Overall, FluentNet shows stronger accuracy and lower miss rates across both datasets and all stutter types, compared to the three variants. Although the drops in performance varies across different stutter types with the removal of each element, the experiment shows the general advantages of the different components of FluentNet. 

The results show that across both datasets, the SE component and the attention mechanism both individually benefit the model for most stutter types. Removal of the SE component yields the greatest drop in the accuracy and increase in miss rates across nearly all stutter types.
The removal of the SE components from FluentNet has the most negative impact. 
The removal of the global attention mechanism as the final stage of the model, also reduces the classification accuracy of FluentNet. 
Similarly, with both the SE component and attention removed, the model 
showed a decline in accuracy and miss rates across all classes tested. Note that the results of these ablation experiments hold similar conclusions for both the UCLASS and our synthesized dataset (with a slightly higher impact observed on UCLASS vs. LibriStutter), thereby reinforcing the validity of LibriStutter's similarity to real stutters. 

\section{Conclusion} \label{Conclusion}
Of the measurable metrics of speech, stuttering continues to be the most difficult to identify as their diversity and uniqueness make them challenging for simple algorithms to model. To this end, we proposed a deep neural network, FluentNet, to accurately classify these disfluencies. FluentNet is an end-to-end deep neural network designed to accurately classify stuttered speech across six different stutter types: sound, word, and phrase repetitions, as well as revisions, interjections, and prolongations. This model uses a Squeeze-and-Excitation residual network to learn effective spectral frame-level speech representations, followed by recurrent bidirectional long short-term memory layers to learn temporal relationships from stuttered speech. A global attention mechanism was then added to focus on salient parts of speech in order to accurately detect the required influences. Through comprehensive experiments, we demonstrate that FluentNet achieves state-of-the-art results on disfluency classification with respect to other works in the area as well as a number of benchmark models on the public UCLASS dataset. Given the lack of sufficient data to facilitate more in-depth research on disfluency detection, we developed a synthetic dataset, LibriStutter, based on the public LibriSpeech dataset. 
Future works may include improving on LibriStutter's realism, which could constitute conducting further research into the physical sound generation of stutters and how they translate to audio signals. 
Whereas this work focuses on the educational and business applications of speech metric analysis, further work may focus towards medical and therapeutic use-cases.

\section*{Acknowledgment}
The authors would like to thank Prof. Jim Hamilton for his support and valuable discussion throughout this work. We also wish to acknowledge Adrienne Nobbe for her consultation towards this project. 

\ifCLASSOPTIONcaptionsoff
  \newpage
\fi

\small
\bibliographystyle{IEEEtran}
\bibliography{refs}

\end{document}